# Laser induced topological cues shape, guide, and anchor human Mesenchymal Stem Cells

*R. Ortiz [a], S. Moreno-Flores [b], I. Quintana [a,c], MdM Vivanco [d], J.R. Sarasua [e], J.L. Toca-Herrera [f]*

[a] *Ultraprecision Processes Unit, Fundación IK4-TEKNIKER, C/Iñaki Goenaga 5, 20600, Eibar, Gipuzcoa, Spain.*

[b] *Former address: Institute for Biophysics, Department of Nanobiotechnology, University of Natural Resources and Life Sciences of Vienna, Muthgasse 11, 1190 Vienna, Austria*

[c] *Micro and Nanoengineering Unit, CIC microGUNE, Goiru Kalea 9, 20500, Arrasate-Mondragón, Gipuzkoa, Spain.*

[d] *Cell Biology & Stem Cells Unit, CIC bioGUNE, Technology Park of Bizkaia, Ed. 801A, 48160 Derio, Spain*

[e] *University of the Basque Country (UPV/EHU), School of Engineering, Department of Mining and Metallurgy Engineering & Materials Science, Alameda de Urquijo s/n, 48013 Bilbao, Spain*

[f] *Institute for Biophysics, Department of Nanobiotechnology, University of Natural Resources and Life Sciences of Vienna, Muthgasse 11, 1190 Vienna, Austria*

Correspondence should be addressed to: iban.quintana@tekniker.es; jr.sarasua@ehu.es; jose.toca-herrera@boku.ac.at

**Abstract**

This report focuses on the effect of the surface topography of the substrate on the behavior of human mesenchymal stem cells from bone marrow (MSCs) before and after co-differentiation into adipocytes and osteoblasts. Picosecond pulsed laser ablation technology was applied to generate different microstructures (microgrooves and microcavities) on poly (L-lactide) (PLLA), where orientation, cell shape and MSCs co-differentiation were investigated. On flat PLLA, the undifferentiated MSCs showed rounded or elongated shapes, the latter being randomly oriented. On PLLA microgrooves however, MSCs adapted their shape to the groove size and direction and occasionally anchored to groove edges. It was found that adipocytes, contrary to osteoblasts, are highly sensitive to topological cues. Adipocytes responded to changes in substrate height and depth, by adapting the intracellular distribution of their lipid vacuoles to these physical constraints. In addition, the modification of PLLA by laser ablation enhanced the adherence of differentiated cells to the substrate. These findings show that picosecond pulsed laser micromachining can be applied to directly manufacture 3D microstructures that guide cell proliferation, control adipocyte morphology and improve the adhesion of bone and fat tissue.

## 1. Introduction

Adult stem cells are the main source for developing future new strategies in regenerative medicine, cell-based therapy, and tissue engineering [1-3]. Proliferation and differentiation of stem cells *in vivo* are regulated by their microenvironment, known as niche, which comprises both cellular components and interacting signals between them [4,5,6]. These niches, in addition to other functions, provide stem cells with physical anchors (by means of adhesion molecules) and regulate the molecular factors that control cell number and fate [5]. Some of these factors are influenced by cell shape, cytoskeletal tension and contractility [7, 8]. Among the biomaterial properties that affect cell behavior, surface topography has great potential to control cell shape and location [9]. Several researchers have observed that microscale and nanoscale topographies in the form of pillars, grooves, pits or pores can induce the differentiation of human mesenchymal stem cells to a certain cell lineage [10, 11, 12]. In this regard, the design of biomaterials with architectures that mimic natural cell microenvironments may be a powerful tool to better understand and manipulate cell function as a strategy for future cell-based therapeutics. In this context, surface microstructuring and micromanufacturing techniques can play an important role in the field of three-dimensional scaffold fabrication, which may enable *in vitro* cells mimic *in vivo* cells, resembling the cellular tridimensional networks and the structural organization of human tissues [13, 14].

Pulsed laser ablation is a well-established micropatterning tool for almost all type of materials [15-17]. This technology represents a very versatile method for 3D micro-structuring that allows direct and fast processing of a wide variety of geometries, including the micromachining of complex structures on flat and non-flat surfaces. Surface modification by laser technologies is thus a promising technique for scaffold fabrication. Liu and collaborators applied for the first time the ultra-short pulsed laser technique for this purpose [18]. They used a femtosecond pulsed laser to generate micro-structured (i.e. holes, grooves and grids) on substrates made of collagen, and studied the growth, adhesion and viability of human fibroblasts and rat bone marrow MSCs on these patterns. To the best of our knowledge only five subsequent reports have applied femtosecond pulsed laser ablation to create 3D microstructures of biocompatible materials for cell culture [19-23]. Despite the high quality structures that a femtosecond pulsed laser is able to generate, some technical limitations, such as the low processing speed and low efficiency, make difficult its application in industry where power scaling is a necessary requirement. In contrast, picosecond pulsed lasers are easier to be implemented in industrial processes due to their better cost-effectiveness and reliability, which has led to an increased presence of these lasers in the industrial market during the last years. However, it has been scarcely used for scaffold fabrication, which has been reported only once before our previous work. Schlie and collaborators applied picosecond pulsed laser ablation on substrates made of silicon to test cell compatibility [24], while previous work from our group has evaluated and applied this technology to the fabrication of three-dimensional biodegradable scaffolds aimed for cell engineering [25]. In that study, the effect of the so-created topographies on the morphology and proliferation of breast cancer cells cultured on Poly-L-Lactide (PLLA) and Polystyrene (PS) was investigated. In this work, we have examined the effect that laser-created microstructures on PLLA substrates have on cell shape, orientation and adipocyte/osteoblasts co-differentiation of human bone marrow mesenchymal stem cells.

## 2. Materials and methods

*2.1 Materials*

Poly-L-Lactide (PLLA) was supplied by Biomer. PLLA sheets of approximately 300 µm of thickness and a degree of crystallinity of 4% (measured with differential scanning calorimetry) were obtained by thermoforming. Cell culture dishes were fabricated by sealing Nylon rings onto PLLA sheets (see Figure 1).

*2.2 3D Microstructuring technique*

Surface micro-structuring of polymer samples was carried out with a picosecond pulse Nd:YVO$_4$ laser (RAPID: Lumera Laser, Germany) integrated in a micromachining workstation by 3D-Micromac. A detailed description of the experimental set-up and the optimization of the micro-structuring process for PLLA can be found in [25, 26].

By means of pulse overlapping different trenches can be fabricated. Trench width and depth were controlled by selecting an appropriate energy (*E*), frequency (*f*) and overlapping distance between pulses (*d*). By using a galvanometric scanner and appropriate control strategies, any desired topography and geometry can be generated on the workpiece. Flat PLLA substrates (Ra = 240 nm) were laser-irradiated applying an energy of 0.9 µJ at a frequency of 100 kHz, and 5 µm of overlapping distance. After this treatment, the average surface roughness (Ra) was increased to 700 nm (Figure 2a). Microgrooves on PLLA (Figure 2b) were obtained by applying an energy of 2.3 µJ at a frequency of 250 kHz, and 2.4 µm of overlapping distance. Microcavities with different geometrical shapes (i.e. square-like, circular, rectangular and elliptical) of areas in the range $(1.25-5) \cdot 10^5$ µm$^2$ were fabricated by overlapping the grooves with a distance of 2 µm at an energy of 1 µJ (Figure c-d show some of these microcavities). The depth of these microcavities has been set to 40 µm, 8 times as large as the cell height, to ensure cell confinement. Dimensions (width and depth) of the different laser-generated topographies were measured by a mechanical stylus profilometer (Dektak 8, Veeco, USA). According to DIN EN ISO 4288:1998, samples profiles of 4 mm length were considered in the measurement of the average surface roughness (Ra).

*2.3 Cell culture*

Human mesenchymal stem cells from bone marrow (MSCs) were provided by Promocell (Germany). Prior to cell culture, PLLA dishes were UV-sterilized for 30 minutes. Cells were cultured in growth medium (Promocell, Germany) at a cell density of 10000 cells/cm$^2$. After cell seeding, the samples were maintained at 37˚C and 5% $CO_2$. In order to observe cell proliferation and morphology on laser patterns, before seeding, cells were stained with NeuroDio, a green-fluorescent cytoplasmic membrane stain (Promokine, Germany). To study cell differentiation, MSCs were seeded in PLLA dishes (30000 cells/cm$^2$), and were left to proliferate until 100% confluence, which corresponds to 24 hours in culture. Then the growth medium was replaced by a differentiation-induction medium, which contained a 1:1 mix of adipogenic and osteogenic induction media (Promocell), and cells were incubated for 2 weeks changing the induction medium every three days. After two weeks, cells were fixed in 4% formaldehyde for 5 minutes at room temperature. Subsequently, cells were stained with Fast Blue RR Salt/Napthol solution (Sigma-Aldrich, Germany) to tag alkaline phosphatase activity (AP staining), an early indicator of cells that undergo osteogenesis, and with Oil Red O solution (Sigma-Aldrich, Germany) to tag the fat deposits (lipid vacuoles) characteristic of adipogenesis. Cell proliferation and differentiation were observed with an inverted fluorescence microscope Nikon Eclipse TE-2000-S (Nikon, Japan) equipped with a Hg vapor lamp equipped, a black/white digital sight DS-Qi1MC, and filter blocks (CFP: EX 436/20, DM 455, BA 480/40; GFP/FITC: EX 480/40, DM 505 and BA 535/50). Bright field experiments are denoted as (BF) in the text.

*2.4. Data analysis*

Cell number, cell coverage and osteoblast:adipocyte differentiation ratio were obtained from microscopy images using the freeware Image J (http://imagej.nih.gov/ij/). All data were expressed as means ± standard deviation. Statistical analysis was carried out using the Student's *t*-test and the values were considered significantly different when $p<0.05$.

## 3. Results

*3.1. Microgrooves manage MSCs shape and form anchorage points.*

Previously, we found that cell alignment on parallel grooves occurred to the greatest extent when pattern density increased: that means narrow grooves and spacing (10 µm wide, and 15 µm spacing) [25]. Therefore, this array of parallel grooves was fabricated to investigate their effect on the shape and proliferation of undifferentiated MSCs. The center of each PLLA dish was divided into four areas, three of which were filled with parallel grooves at different directions: 0°, 45°, and 90°. A fourth square was left untreated as control (flat PLLA). Laser-created grooves exhibited two types of topographies: depressions with dimensions close to cell size; and protrusions formed at groove edges by the recast material. PLLA recast is particularly prominent at these regions as a consequence of the *first-pulse effect* of the laser (generation of an intense first light pulse) [27].

Figure 3 shows undifferentiated MSCs proliferating on flat (Figure 3a-c) and on PLLA grooves (Figure 3d-i) after 1, 8 and 22 days in culture. On flat PLLA, after one day in culture, MSCs adopted various shapes, with a mixture of rounded and elongated in different directions. After twenty-two days and 80% cell confluency, cell bundles formed with a particular orientation, maybe due to enhanced cell-cell interactions. In contrast, on PLLA grooves, MSCs with elongated shapes predominate already after 1 day in culture. The direction of these elongated cells coincides with the groove orientation as Figure 3g-i show, which clearly confirms the influence of substrate topography on the early alignment of cells.

The effects of substrate topography on cell orientation were quantified. Figures 4a-c depict polar plots of the number of cells with elongated shapes, normalized to the total cell number, as a function of groove orientation on flat PLLA and PLLA grooves after 1, 8 and 22 days in culture. Cells with rounded shapes were considered as non-oriented. The graphs show that at low cell confluence the number of oriented cells and angle of orientation remain constant with time in culture, regardless of the PLLA topography (i.e., flat PLLA or grooves oriented at 0° and 45°).

Figure 4d represents the number of oriented cells normalized to the total cell number as a function of time in culture. At times as long as 8 days, cell alignment was clearly more frequent on PLLA grooves than on flat PLLA. After 22 days, cell alignment on flat PLLA increased to values comparable to those obtained on PLLA grooves, however this increment is not statistically significant ($p>0.3$, according to the Student t-Test) due to the high associated error (20%). On PLLA grooves, cell alignment remained constant at a value of 70% (± 10%).

Figure 5 shows undifferentiated MSCs in the gaps between grooved areas, anchored to the protrusions at the groove edges after 1 day in culture. As the images indicate, cells readily spread across the 200 µm wide interspace between patterned areas by developing filopodia-like protrusions that seem to anchor to groove edges

*3.2. Effect of surface topography on co-differentiation of human mesenchymal stem cells*

In-vitro differentiation of MSCs is most efficient when it is induced at high cell confluency (i.e. 90% [28]). Under these conditions, it has been shown that cell-cell interactions may be more relevant than cell-substrate interactions to determine the shape and orientation of MSCs [28]. Nevertheless, when a micropatterned surface is considered, changes on the surface roughness as well as the presence of topological barriers and cavities altered the behaviour of differentiated cells as we will show in the following sections.

*3.2.1. Topological barriers alter the intracellular distribution of lipid vacuoles*

Descending from MSC precursors, differentiated adipocytes are round to allow for maximal lipid storage in adipose tissue, while osteoblasts tend to spread to facilitate matrix deposition activity. To determine the level of adipogenic differentiation, the distribution of lipid vacuoles cultured on flat or rough PLLA for 14 days was examined. The clusters of lipid vacuoles were almost rounded and their distribution was not affected by surface roughness on PLLA (Figure 6a-b). Similarly, AP staining suggested that osteoblasts distribution was unaffected. However the lipid vacuoles showed an aligned distribution along the edge between the two zones of different roughness (Figure 6c). Similar behavior was observed on the edges of laser-microcavities, where cells found a topological barrier of approximately 40 µm high (Figures 6d-e). Lipid vacuoles close to the edge of these cavities lined up along the borderline irrespectively of the size of the microcavity, which in this case was much larger than a single cell.

In contrast, on PLLA grooves, where two topological barriers are just a few micrometers apart (i.e. 15 µm), the distribution of the lipid vacuoles was different (Figures 7-c). Lipid vacuoles arranged in strings both inside (67 ± 14 %) and in between (29 ± 13 %) the grooves (Figure 7d). In contrast, substrate topology did not induce any noticeable effect on the osteoblast shape.

*3.2.2. Laser surface modification enhance the adherence of differentiated human mesenchymal stem cells*

Laser treatment alters surface roughness, the extent of which mainly depends on the parameters that control laser performance. In our case, laser irradiation increased the roughness of PLLA by 3-fold. We would like to determine whether the resulting surface modification may in turn affect the extent of differentiation of MSCs, the adipocyte:osteoblast differentiation ratio (i.e. codifferentiation) or both.

To analyze further the potential effects of the different substrates on MSC differentiation, the area covered by each cell type was determined. The histogram in Figure 8 shows the percentage of area covered by undifferentiated MSCs, cells differentiated into adipocytes, and cells differentiated into osteoblasts on different substrates: flat PLLA (control), laser-irradiated (rough)

PLLA and laser microstructured PLLA (grooves). The area covered by MSCs that have undergone differentiation (approximately 90%) remained constant irrespectively of the type of substrate ($p$>0.1, according to the Student $t$-Test). Regarding the adipocyte:osteoblast differentiation ratio, it was observed that, on all types of substrate, the area covered by osteoblasts was larger than that covered by adipocytes.

While laser treatment thus appears not to have an effect on the co-differentiation of MSCs, it certainly did affect cell adherence upon differentiation as shown in Figure 9a-c. MSCs were seeded on partially laser-treated PLLA dishes at high cell confluence to form a continuous layer and allowed to co-differentiate for as long as 14 days. During differentiation, part of the cell layer spontaneously detached from the substrate. The low-magnification micrographs show that only those cells cultured on laser-treated areas remained adhered, indicating that the cell layer actually detached from the untreated (flat) PLLA surface. The enhancement of cell-substrate interactions may actually occur on laser-irradiated areas, regardless of the presence of microstructures (Figures 9a-b) or microsized barriers (such as in cavities, Figure 9c).

**4. Discussion**

We have examined the effects that laser-created microstructures on PLLA have on MSC shape, orientation and differentiation. MSCs cultured on PLLA micro-grooves showed the cell contact guidance effect [29] already 24 h after seeding, and it was still noticeable 22 days later. Our results agree with previous reports on the alignment of stem cells along micro- and nano-grooved-patterned substrates [30-32, 23]. In addition, the protrusions generated by our technique at these points seemed to favor cell anchoring and elongation between them. Hamilton and collaborators observed a similar phenomenon in osteoblasts proliferating on boxes and pillars [33], which they called gap guidance, a type of contact guidance for cell alignment that is associated to discontinuous topographical edges. Our results on microgrooves show that these microstructures influence MSCs in two ways: on the one hand, grooves influence cell orientation, since cells adapted their shape to groove width and orientation (the maximal effect was observed at the first stages of MSCs proliferation); on the other hand, the grooves promote cell adherence and guidance at their edges. These effects appear to resemble the influence of stem cell niches *in* *vivo* [5]. Further experiments would be necessary to examine the extent of the gap guidance, the topology of the edge protrusion and its spatial distribution that altogether make groove edges effective anchor points to MSCs and guide their shape

It was found that adipocytes, in contrast to osteoblasts, are highly sensitive to surface topography. They responded to every type of laser surface modification, adapting their morphology to the created topographies. Adipocyte differentiation on grooved PLLA showed the alignment and confinement of the lipid vacuoles in strings along the grooves. Similarly, adipocytes adapted the distribution of lipid vacuoles to follow the borderline between flat and rough PLLA, and also the 40 µm-deep edge of the microcavities. To the best of our knowledge, this is the first observation of adipocyte compliance to topological cues. The findings suggest that surface topography is a potential tool to control (or modify) adipocyte morphology. Mature adipocytes release a big variety of factors that play a fundamental role in the regulation of many important functions of the body, such as the regulation of appetite, energy homeostasis, insulin sensitivity, immunological responses and vascular diseases [34]. The expression of these factors is, in many cases, controlled by adipocyte size and location [35, 36]. In this context, adipocytes should be able to change their morphology in order to store the optimum amount of fat and to perform properly their physiological functions [37]. Therefore, it is likely that if surface topography influences adipocyte morphology, it could also affect the expression of different factors. This may have beneficial consequences in nano- and regenerative medicine, since surface topography could be tailored to treat diseases related to the functions regulated by these factors. Additionally, specially designed scaffolds can be developed for the formation of adipose tissue [38].

The effect of physical constraints on cell behavior and differentiation has been investigated in this work. Contrary to results reported on chemically confined cells (such as adhesive/non-adhesive treatments) [7, 39-41], in our study, there were not significant differences between cell differentiation inside and outside the microcavities of different shapes and at these dimensions. However, it is relevant to point out that our experiments showed some other differences compared to the studies conducted by other authors. Under the present conditions, where no chemical restrictions were imposed on the cells, we did not induce mechanical force gradients in the borderline of the microcavities, which can explain the observed differences.

Cell adherence of differentiated cells is affected by the substrate. When cells differentiate into adipocytes and osteoblasts on flat PLLA, the cell monolayer detached easily from the substrate; this suggests cell-cell interactions outperform cell-substrate interactions during differentiation, which may be related to the formation of extracellular matrix (ECM) [42]. However, the cell monolayer remained anchored to the substrate on laser-roughened PLLA, grooves and microcavities, suggesting that laser-induced roughness in every type of microstructure enhances cell-substrate interactions regardless of shape or dimensions, and without significantly affecting cell type differentiation. Therefore, laser surface modification could be applied to directly improve cell adhesion without producing adverse effects on material biocompatibility, as observed in other tissue adhesives [43, 44].

## 5. Summary and Conclusions

The present study shows that topographical features at micrometer scale are able to influence the morphology and orientation of human mesenchymal stem cell and adipocytes, without the interplay of chemical factors. Laser-generated microstructures induced the cell contact guidance effect on human mesenchymal stem cells, favoring cell organization, directing cell anchorage, and enhancing cell adherence. In addition, adipocytes, contrary to osteoblasts, are highly sensitive to substrate microtopography, and accommodate along topological edges and barriers at different degrees of confinement. Furthermore, laser modification on PLLA enhances the adherence of the cell monolayer during osteogenesis and adipogenesis differentiation without compromising the latter. In view of these results, laser microstructuring represents a potential tool for providing a 3D cell microenvironment representative of stem cell niches in vivo, enhancing cell adherence and organization. Although it is likely that both physical and chemical surface modification is required to control cell differentiation and induce cell fate, our experimental approach could contribute to enhance the biocompatibility of implants, as well as to improve existing surgical procedures that involve healing of soft tissue.

**Acknowledgements**

This work was partially funded by the Basque Government - (UE09+/13) and the IGS BioNanoTech Program of the Federal Ministry for Science and Research, Austria.

**Figure captions**

Figure 1. Typical PLLA-bottom dish used for these studies. The thermoformed PLLA sheet (approximately 300 µm thick) is sealed to a Nylon ring with a biocompatible silicone (blue-green, Picodent Twinsil, Picodent GmbH, Germany).

Figure 2. SEM-obtained images of laser-created topographies: roughened PLLA (a); grooves (10 µm wide, 7 µm deep and 15 µm spacing) (b); and circular and square-shaped microcavities (c-d).

Figure 3. Fluorescence microscopy images of undifferentiated MSCs cultured on PLLA grooves. Images a-i show cells on flat PLLA (a-c) and PLLA grooves (d-i) as a function of time. Images (g-i) show cells on grooves oriented at different directions after one day in culture.

Figure 4. Polar diagrams showing the number of oriented cells as a function of the orientation angle after 1 (a), 8 (b) and 22 (c) days in culture. Size of dots indicates the error. (d) Population histogram of oriented cells normalized to total cell number obtained from fluorescence images as a function of time.

Figure 5. Undifferentiated MSCs between three (a) and two (b) grooved-patterned areas were visualized by inmunofluorescence after 1 day in culture.

Figure 6. Brightfield (BF) images of MSCs differentiated into adipocytes (presence of lipid vacuoles in red) and osteoblasts (AP staining in blue) on flat PLLA (a), laser-roughened PLLA (b,c), and microcavities (d, e). In figures c-e strings of lipid vacuoles are outlined.

Figure 7. BF images of MSCs differentiated into adipocytes (presence of lipid vacuoles in red) and osteoblasts (AP staining in blue) on PLLA grooves at different directions (0°, 45°, 90°) after 14 days in culture (a-c). (d) Quantification of the proportion of clusters of lipid vacuoles confined in the grooves and on the intergroove space (*significance level according to the Student t-Test: $p<0.005$).

Figure 8. Histogram of the area (represented as percentage) covered by each cell type (adipocytes, osteoblasts, and undifferentiated MSCs) on control (flat PLLA), rough PLLA, and PLLA grooves.

Figure 9. Monolayers of differentiated MSCs remain adhered only on laser-treated areas: rough PLLA (a), PLLA grooves (b), or microcavities (c). Micrographs taken after 14 days in culture.

Figure 1

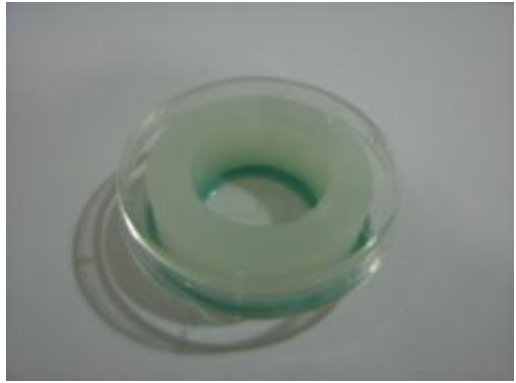

Figure 2

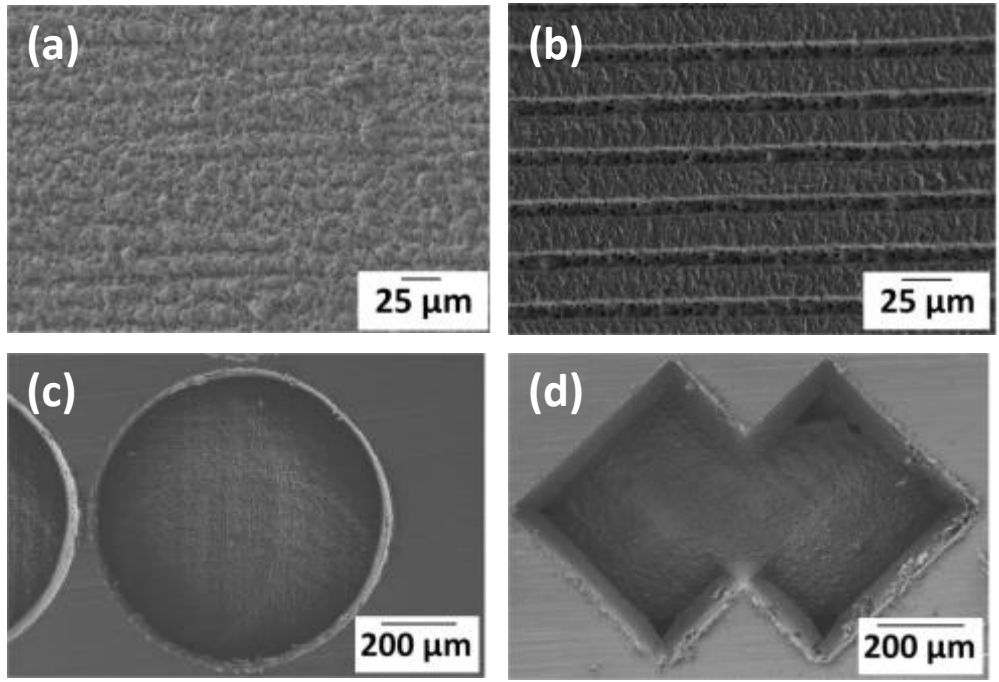

Figure 3

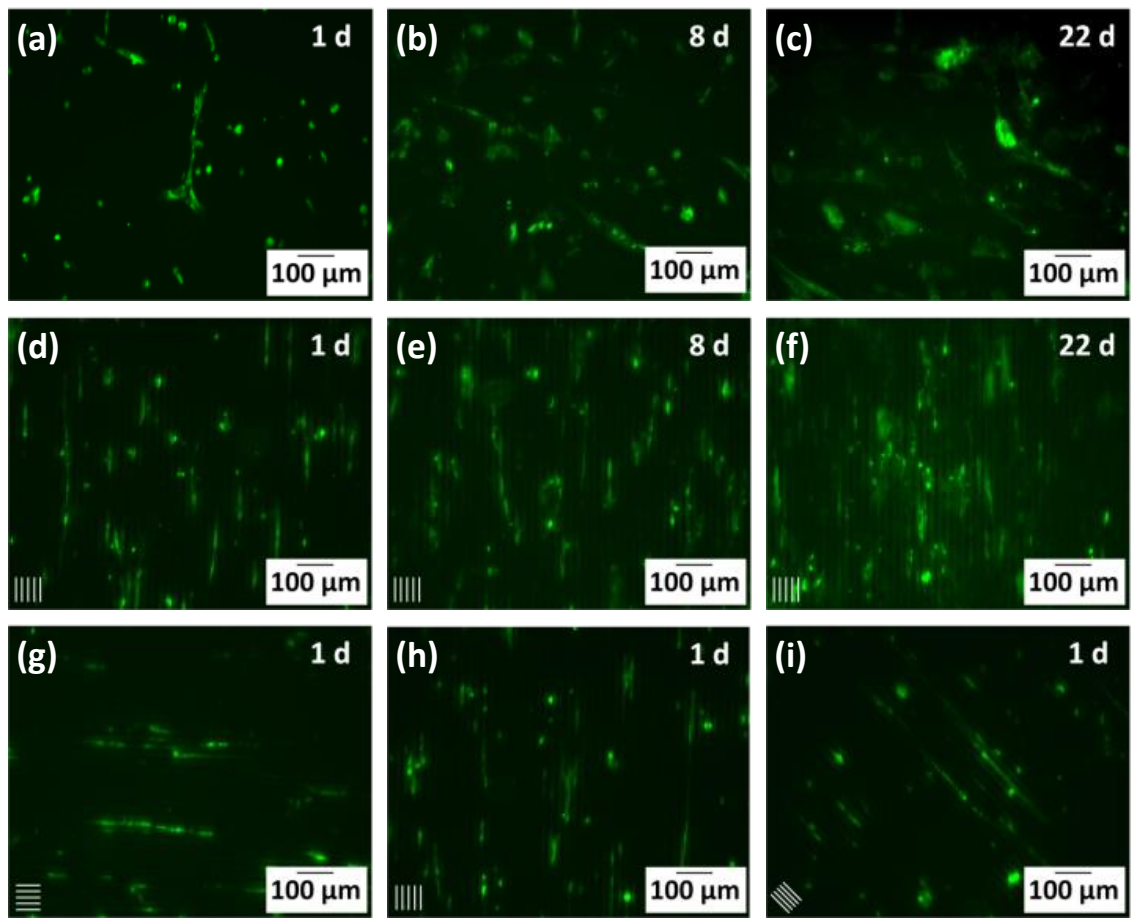

Figure 4

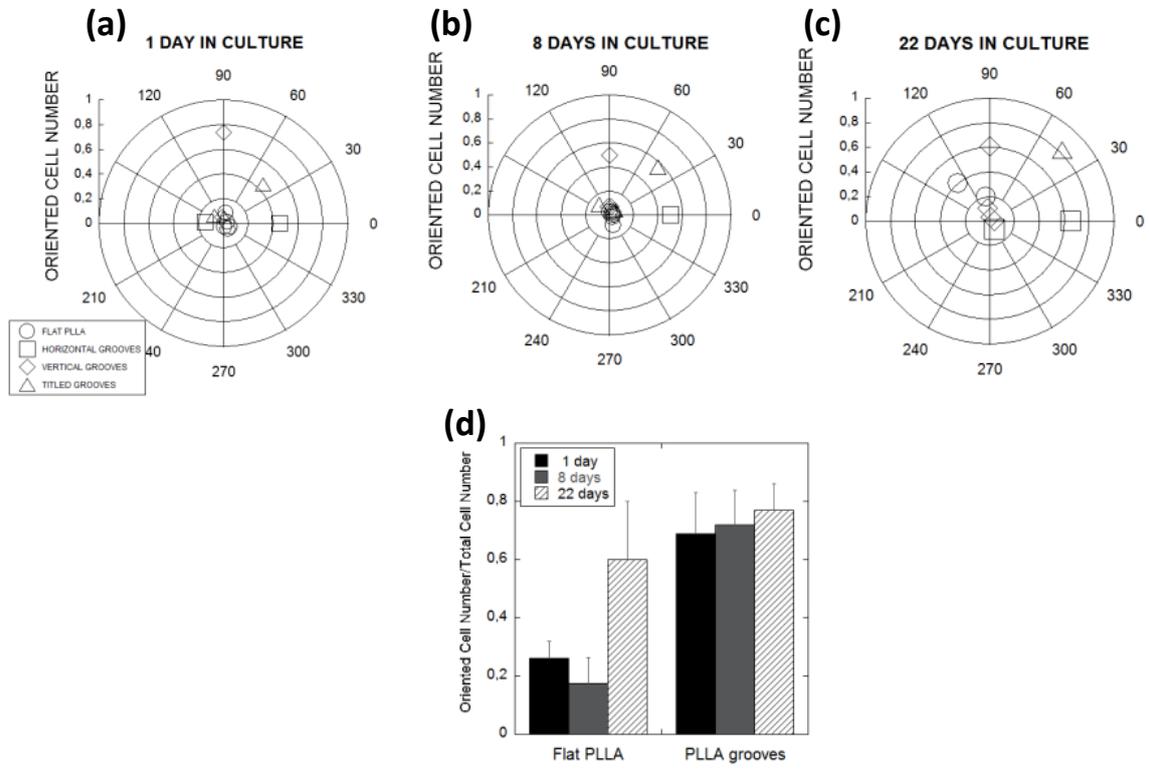

Figure 5

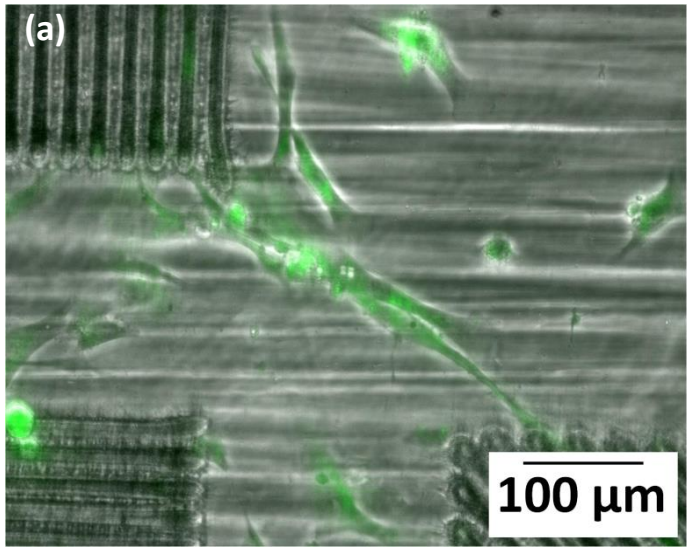

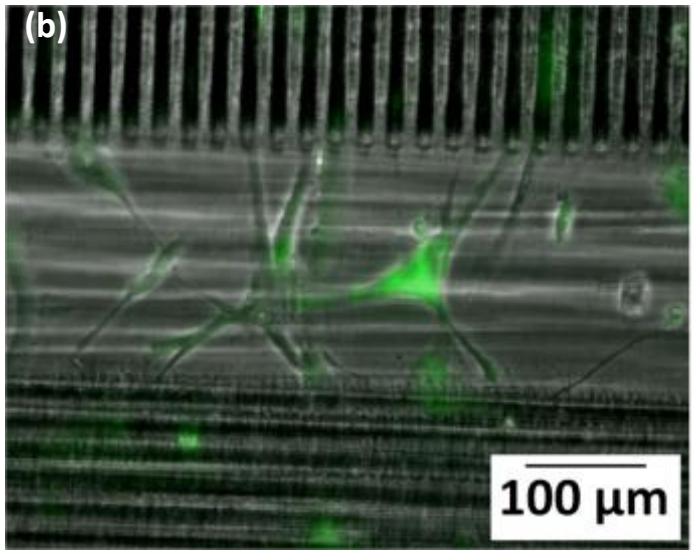

Figure 6

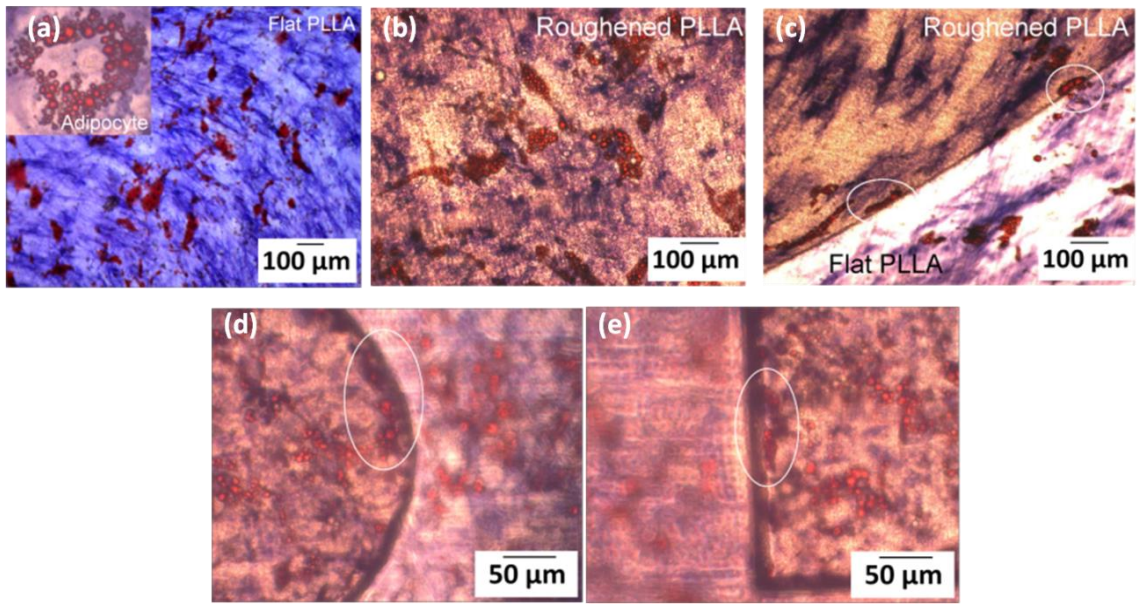

Figure 7

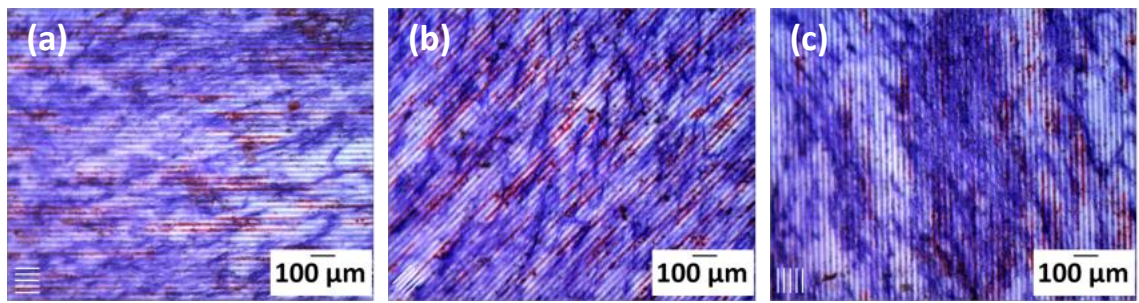

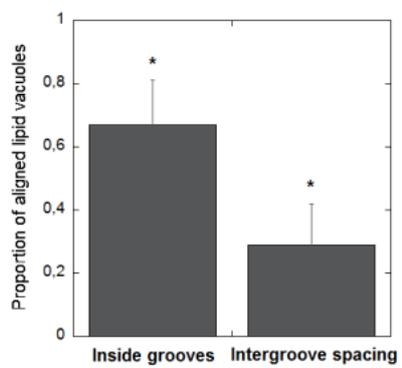

Figure 8

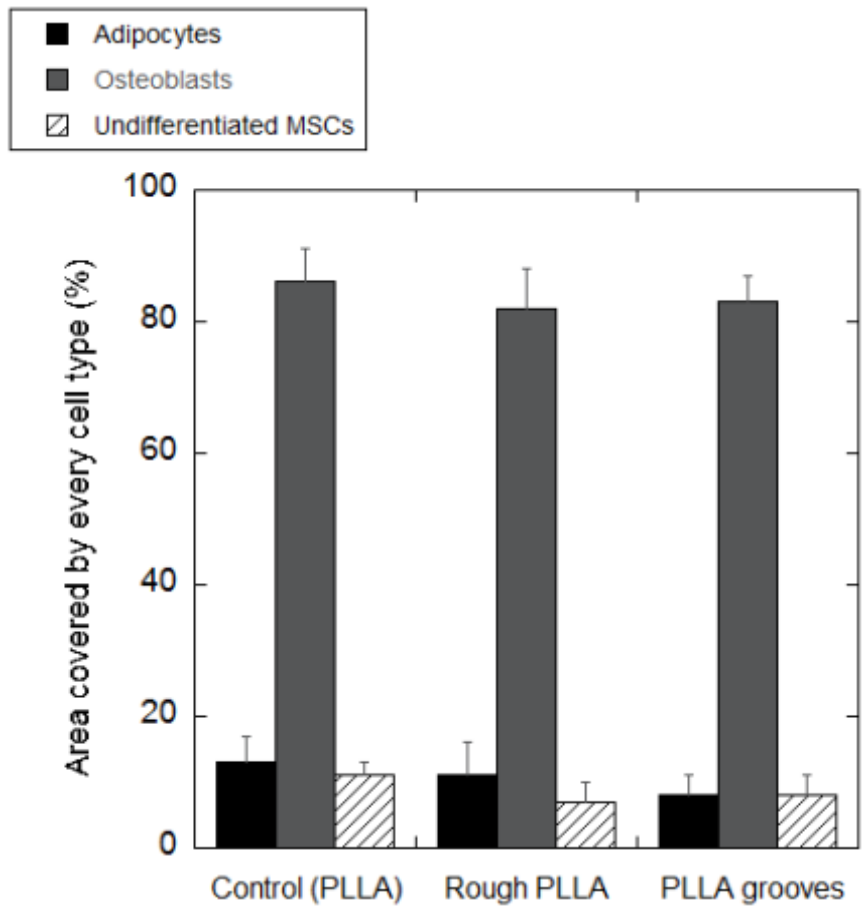

Figure 9

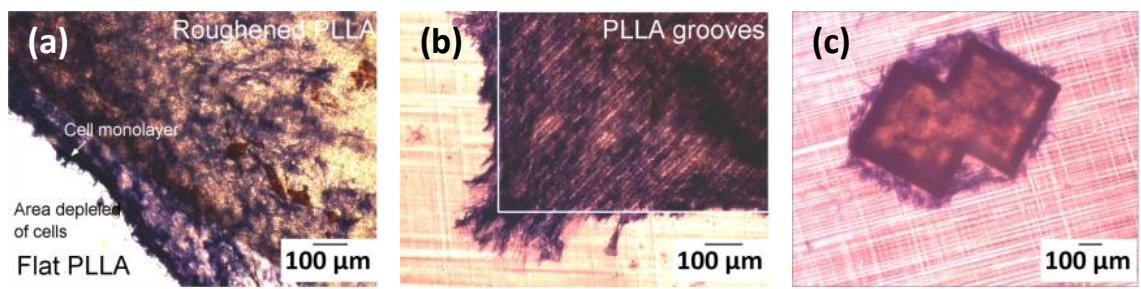